\documentclass[aps,twocolumn,prl, grouppedaddress, longbibliography]{revtex4-1}
\usepackage{amsmath,amssymb}
\usepackage{dsfont,yfonts}
\usepackage{pgfplotstable}
\usepackage{pgfplots}
\usepackage{tabularx}
\usepackage{multirow}
\usepackage{diagbox}
\newcolumntype{C}{>{\centering\arraybackslash}X}%
\usepackage{graphicx}
\usepackage{subfigure}
\usepackage{tikz}
\usepackage{pgffor}
\usepackage{verbatim}
\usepackage{float}
\usepackage{mathrsfs}
\usepackage[colorlinks, bookmarks=true,breaklinks=true,linkcolor=red, citecolor=blue, linktocpage=true, urlcolor=blue]{hyperref}

\newcommand{\bfk}{\mathbf{k}}
\newcommand{\bfp}{\mathbf{p}}

\makeatletter
\newcommand\xleftrightarrow[2][]{%
  \ext@arrow 9999{\longleftrightarrowfill@}{#1}{#2}}
\newcommand\longleftrightarrowfill@{%
  \arrowfill@\leftarrow\relbar\rightarrow}
\makeatother

\begin{document}

\title{Thermodynamically Induced Transport Anomaly in Dilute Metals ZrTe$_5$ and HfTe$_5$}

\author{Chenjie Wang}
\email{cjwang@hku.hk}
\affiliation{Department of Physics and HKU-UCAS Joint Institute for Theoretical and Computational Physics, The University of Hong Kong, Pokfulam Road, Hong Kong, China}

\date{\today}

\begin{abstract}
A 40-year-old puzzle in transition metal pentatellurides ZrTe$_5$ and HfTe$_5$ is the anomalous peak in the temperature dependence of the longitudinal resistivity, which is accompanied by sign reverses of the Hall and Seebeck coefficients. We give a plausible explanation for these phenomena without assuming any phase transition or strong interaction effect.  We show that due to intrinsic thermodynamics and diluteness of the conducting electrons in these materials, the chemical potential displays a strong dependence on the temperature and magnetic field. With that, we  compute resistivity, Hall and Seebeck coefficients in zero field, and magnetoresistivity and Hall resistivity in finite magnetic fields, in all of which we reproduce the main features that are observed in experiments.
\end{abstract}

\maketitle

\emph{Introduction.} A 40-year-old puzzle in transition metal pentatellurides ZrTe$_5$ and HfTe$_5$ is the anomalous peak in the temperature dependence of resistivity and the accompanying sign reverses of Hall and Seebeck coefficients\cite{AnomalousPeak1,AnomalousPeak2, HallSignReverse1, HallSignReverse2, HallSignReverse3}. The peak temperature $T_p$ varies in the range $0\sim200$K in different samples. Early attempts to explain these transport anomalies through a structural transition or charge/spin density waves failed\cite{DiSalvoPRB1981,OkadaJPSJ1982}. A recent theoretical proposal\cite{WengPRX2014} that they are good candidates of topological insulators or Dirac/Weyl semimetals has motivated a great effort to reinvestigate the two materials, leading to many interesting discoveries such as the chiral magnetic effect\cite{LiNatPhys2016} and 3D quantum Hall effects\cite{ZhangNature2019}. Regarding the puzzle, important progress was made by ARPES experiments\cite{ZhangNatComm2017,ZhangSciBull2017} (see also Refs.~\cite{WuPRX2016, MoreschiniPRB2016,ChiNJP2017}): it was observed that as temperature increases, the chemical potential shifted from the electron-like conduction band to the hole-like valence band, consistent with the change of charge carrier type. This observation was attributed to temperature-induced Lifshitz transition, but the underlying reason remains unclear. Other explanations to the puzzle are also proposed, such as polaronic models with strong electron-phonon coupling\cite{RubinsteinPRB1999,FuPRL2020}, bipolar conduction\cite{ShahiPRX2018}, semimetal-semiconductor transition\cite{McIlroyJPCM2004} or topological phase transition\cite{ZhaoCPL2017,XuPRL2018}. So far, the problem is still under debate.

In this paper, we show that the above puzzle can be well resolved by intrinsic thermodynamics of non-interacting electrons, without the need of a phase transition or strong interaction. One of our key observations is that there are two small energies in ZrTe$_5$/HfTe$_5$: the Fermi energy $E_F$ and a particle-hole (PH) symmetry breaking energy $\Delta$. The latter implies that the density of states (DOS) becomes very asymmetric between the conduction and valence bands for energy bigger than $\Delta$. Estimates from experiments are $E_F=15\sim40$meV and $\Delta=30\sim40$meV. We show that the smallness of $E_F$ and $\Delta$ makes the chemical potential highly sensitive to the temperature $T$ and external magnetic field $B$, leading to the experimentally observed chemical potential shift\cite{ZhangNatComm2017, ZhangSciBull2017}. With this thermodynamic property and the Kubo formula, we are able to reproduce the main transport features observed in experiments, including the resistivity peak and sign reverses of the Hall and Seebeck coefficients.

\begin{figure}[b]
\centering
\includegraphics[scale=1]{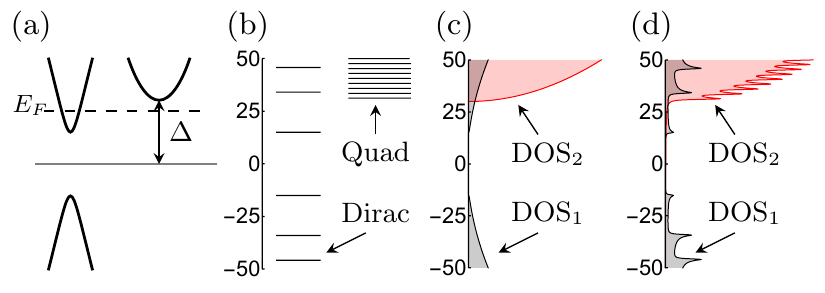}
\caption{(a) Schematics of the low-energy band structure in ZrTe$_5$. (b) Landau level bottoms of 3D Dirac and quadratic fermions with $B=4$T (energy in units of meV; see Fig.~\ref{fig:muT} for numerics).  (c) Densities of states at $B=0$ and (d) at $B=4$T (smoothed by a small disorder).}
\label{fig:dos}
\end{figure}

\begin{figure*}
\centering
\includegraphics[scale=1]{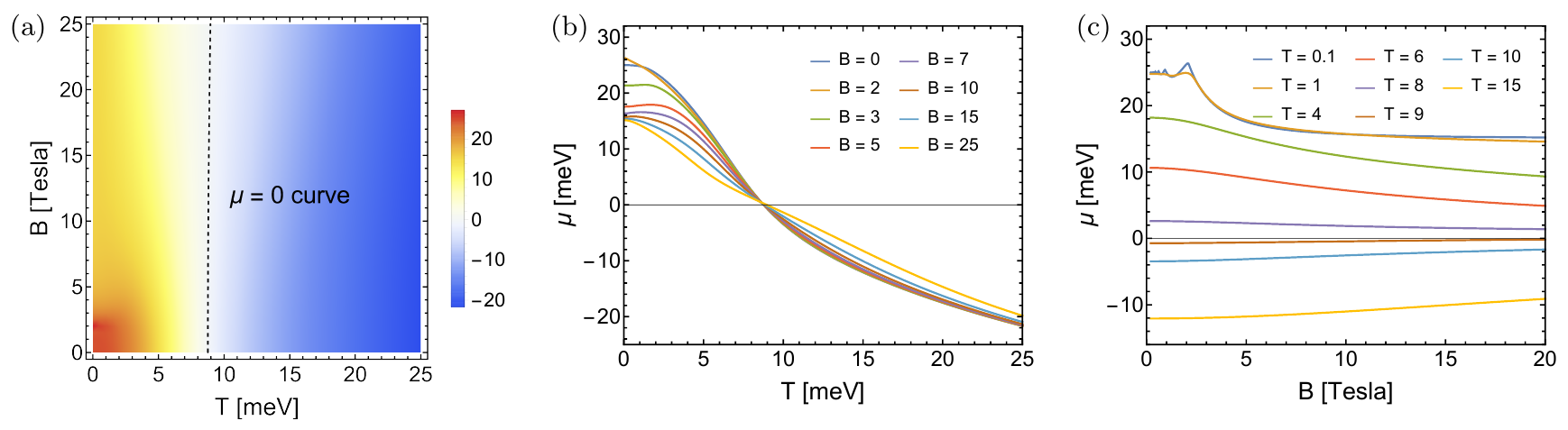}
\caption{Plots of the chemical potential $\mu$ versus temperature $T$ and magnetic field $B$, obtained from Eq.~\eqref{eq:density} with parameters set approximately to the experimental values in Refs.~\onlinecite{ZhangNature2019,ZhangNatComm2017}: $E_F=25$ meV, $m=15$ meV,  $\Delta=30$ meV, $\hbar v_x = 6.0$ eV$\cdot$\AA, $\hbar v_y = 1.3$ eV$\cdot$\AA, $\hbar v_z = 0.2$ eV$\cdot$\AA, $m_x^*=m_y^*=0.2m_e$, $m_z^*=2m_e$ with $m_e$ being the electron mass.  The effective masses $m_x^*, m_y^*, m_z^*$ of the quadratic fermions have not been measured experimentally to our knowledge, so they are set to typical values with anisotropy taken into account. The associated number density is $n=1.73\times 10^{17}$ cm$^{-3}$, and cyclotron energies are $\hbar\omega_{c1}= 15.3\sqrt{B}$ meV and $\hbar\omega_{c2}=2.4B$ meV. Numerical data in other figures are the same as here if not otherwise specified.}
\label{fig:muT}
\end{figure*}

\emph{Model.} Our discussions will focus on ZrTe$_5$ but they can be easily adapted to HfTe$_5$. ZrTe$_5$ is a highly anisotropic layered material. According to Ref.~\onlinecite{ZhangNatComm2017}, the low-energy band structure contains a Dirac-like electron pocket at $\Gamma$ point and four other electron pockets near the Brillouin zone boundary. We model them by an anisotropic Dirac fermion and four identical anisotropic quadratic fermions[Fig.~\ref{fig:dos}(a)]. The Dirac fermion has the well-known relativistic dispersion $E = \pm \sqrt{m^2 + v_x^2p_x^2 +v_y^2p_y^2+v_z^2p_z^2 }$, where $m$ is the Dirac mass and $v_\alpha$ is the velocity in $\alpha$ direction, $\alpha=x,y,z$. The energy bottom of the quadratic fermions is $\Delta$, measured from the midpoint of the Dirac dispersion. 
For the Dirac and quadratic fermions, densities of states per volume are given, respectively, by
\begin{align}
D_1(\epsilon) & = 2\alpha_1 |\epsilon| \sqrt{\epsilon^2-m^2} \  \Theta(|\epsilon|-m), \nonumber\\
D_2(\epsilon) & = 2\kappa \alpha_2 \sqrt{\epsilon-\Delta} \ \Theta(\epsilon-\Delta),
\label{eq:dos}
\end{align}
where  $\alpha_1 =1/(2\pi^2\hbar^3v_xv_yv_z)$, $\Theta(x)$ is the Heaviside step function, $\kappa=4$ denotes the four copies of quadratic fermions,  $\alpha_2 =\sqrt{2m^*_xm_y^*m_z^*} /(2\pi^2\hbar^3)$, $m_\alpha^*$ is the anisotropic effective mass of quadratic fermions, and the factor 2 comes from spin degeneracy.  The total DOS is $D(\epsilon)=D_1(\epsilon)+D_2(\epsilon)$. We emphasize that the Dirac dispersion is PH symmetric, i.e., $D_1(\epsilon) = D_1(-\epsilon)$. The presence of the quadratic fermions breaks the PH symmetry.

We also consider the effect of an external magnetic field $\mathbf{B}=B\hat{z}$, under which electron eigenstates form Landau levels. Details on Landau levels of the Dirac fermion can be found, e.g., in Refs.~\onlinecite{NetoRMP2009, ShenBook} or Supplemental Materials (SM).\footnote{The Supplementary Material contains technical analyses on $\mu(T,B)$,  methods for computing transport coefficients, and additional discussion on zero-field transport, where Refs.\cite{LiangNatPhys2018,ZhangNatComm2021} are cited.} 
The Landau level energy is given by $E_a=\lambda\sqrt{m^2 + \hbar^2\omega_{c1}^2 N + v_z^2p_z^2}$, where $N\ge 0$ is the Landau level index, $p_z$ is the momentum along $z$ direction, $\lambda=\pm 1$ represents the electron or hole branch respectively, $\omega_{c1} = \sqrt{2v_xv_yeB/\hbar c}$ is relativistic cyclotron frequency and Zeeman splitting is neglected.  Landau levels of quadratic fermions are textbook results, with the energy $E = \hbar\omega_{c2}(N+1/2) + p_z^2/2m_z^* + \Delta$, where $\omega_{c2}= eB/(c\sqrt{m_x^*m_y^*})$. The densities of states are now given by
\begin{align}
D_1(\epsilon,B) & = \alpha_1 \sum_{N\ge 0}  \frac{d_N\hbar^2 \omega_{c1}^2 |\epsilon|}{2\sqrt{\epsilon^2-E_{N1}^2}} \Theta(|\epsilon|-E_{N1}) \nonumber\\
D_2(\epsilon,B) & = \kappa\alpha_2 \sum_{N\ge0 } \frac{\hbar \omega_{c2}}{\sqrt{\epsilon-E_{N2}}}\Theta(\epsilon-E_{N2})
\label{eq:dos-B}
\end{align}
where $d_N=2-\delta_{N,0}$, $E_{N1} = \sqrt{m^2 + \hbar^2\omega_{c1}^2 N}$, and $E_{N2} = \hbar\omega_{c2}(N+1/2)+\Delta$. The latter two are the bottom energies of 3D Landau levels of the Dirac and quadratic fermions respectively. In the limit $B\rightarrow 0$, expressions in \eqref{eq:dos-B} reduce to those in \eqref{eq:dos}.

Figure~\ref{fig:dos}(b) plots the Landau level bottoms at $B=4$T to give readers a sense of level spacings, with numerics given in the caption of Fig.~\ref{fig:muT}. We note that the Dirac fermion is  ``lighter'' than the quadratic fermions, and thereby has bigger level spacings. Figures \ref{fig:dos}(c) and \ref{fig:dos}(d) show  densities of states in zero field and in a finite $B$ field respectively. Effect of disorder is neglected in our thermodynamic calculations. It will be included in our calculations of transport properties below.

\emph{Chemical potential $\mu(T,B)$.} With the above densities of states, we now study the $T$ and $B$ dependence of the chemical potential $\mu$. The particle number density $n(T,B,\mu)$ can be expressed as 
\begin{align}
n(T,B,\mu) = & \int_{0}^\infty\!\!\! d\epsilon D(\epsilon,B)f_{T}(\epsilon-\mu)  \nonumber\\
&  + \int_{-\infty}^0\!\!\! d\epsilon D(\epsilon,B)[f_T(\epsilon-\mu)-1], \label{eq:density}
\end{align}
where $f_T(\epsilon) = 1/[\exp(\epsilon/T)+1]$ is the Fermi-Dirac distribution (Boltzmann constant $k_B$ is absorbed into $T$ throughout the paper). Charge neutrality is taken to be at $\epsilon=0$. The density $n$ is fixed in a given 3D sample, so Eq.~\eqref{eq:density} should be understood as an integral equation that defines a function $\mu(T,B)$.  

We solve Eq.~\eqref{eq:density} numerically by setting all the parameters in \eqref{eq:dos} and \eqref{eq:dos-B} to be comparable to experimentally measured values.\cite{ZhangNature2019}   The results are shown in Fig.~\ref{fig:muT}.  When $T$ varies in the range $0\sim 25$meV and $B$ varies in the range $0\sim 25$T, our calculation shows a significant variation in $\mu$, of the order of $E_F$, in agreement with the ARPES observation\cite{ZhangNatComm2017}. In particular, for $B=0$, the chemical potential $\mu$ decreases monotonically from a positive $E_F$ into a negative value around $T_0\approx 10$meV. One can show that this monotonic decreasing behavior occurs for a fairly general class of densities of states --- see SM\cite{Note1}. A monotonic behavior is also observed when $T \gtrsim \hbar\omega_{c1}$ for a finite $B$. When $T\lesssim \hbar\omega_{c1}$, quantum  oscillations in $\mu$ can also be seen in Fig.~\ref{fig:muT}(c), which, however, is not our focus.

The fact that $\mu$ changes so dramatically, compared to conventional metals, in the temperature regime $T \lesssim 25$meV and in experimentally accessible magnetic fields $B\lesssim 25$T follows from two properties of ZrTe$_5$: (i) $E_F$ is small, approximately $25$meV,  i.e., conducting electrons are dilute; and (ii) the PH symmetry is broken at the energy scale of $\Delta$, and $\Delta \approx E_F$. Generally speaking, variation of $\mu$ is set by $T/E_F$ and $\hbar\omega_{c1}/E_F$. Accordingly, a small $E_F$ makes it easier to achieve a significant change in $\mu$ by tuning $T$ and $B$. Nevertheless, without property (ii),  one can show that the PH symmetry guarantees  $\mu> 0$, if $E_F>0$. Or equivalently,  PH symmetry pushes the sign-reversing temperature $T_0$, defined by $\mu(T_0,B)=0$, to infinity. To have a finite $T_0$, the PH symmetry must be broken. We will show below that $T_0$ is closely related to the sign-reversing temperature of the Hall and Seebeck coefficients. While the scale of $T_0$ is set by $\Delta$, its precise value depends on $E_F$, the effective masses $m_\alpha^*$, magnetic field $B$, etc.  We note that in Fig.~\ref{fig:muT}(b), $T_0$ barely displays a dependence on $B$. However, there is actually a very weak quadratic correction, $\delta T_0\propto B^2$, which we discuss in SM\cite{Note1}. We remark that a nonzero Dirac mass $m$ is not important. Very similar behaviors of $\mu$ and transport properties are obtained for $m=0$ (see SM\cite{Note1} for more discussions).

\begin{figure}
\centering
\includegraphics[scale=1]{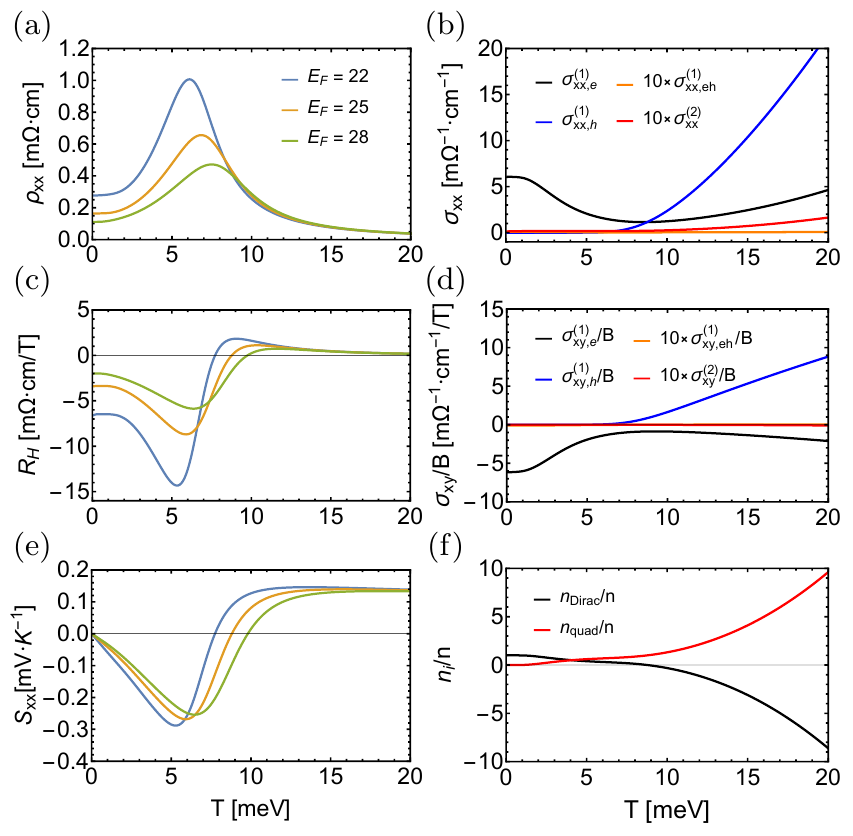}
\caption{Temperature dependence of the longitudinal resistivity $\rho_{xx}$ (a), Hall coefficient $R_H$ (c) and Seebeck coefficient $S_{xx}$ (e) with different Fermi energies.  The level broadening constants are set by $\Gamma_1=0.5$ meV (estimated from experimental data in Ref.~\onlinecite{ZhangNature2019}) and $\Gamma_2=10\Gamma_1$ (see an estimate in SM\cite{Note1} using Born approximation) throughout our calculations. (b), (d) Different contributions to the longitudinal conductivity $\sigma_{xx}$ and Hall conductivity $\sigma_{xy}$ at $E_F=25$meV. (f) The density ratios $n_{\rm Dirac}/n$ and $n_{\rm quad}/n$ versus temperature.} 
\label{fig:transport_B0}
\end{figure}

\begin{figure*}
\centering
\includegraphics[scale=1]{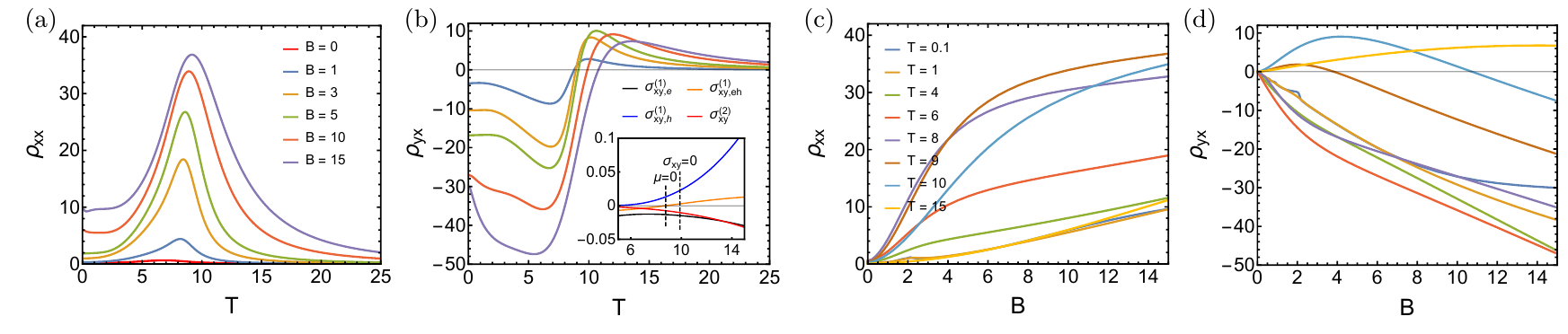}
\caption{Calculated magnetoresistivity and Hall resistivity. (a), (b) Temperature dependences of magnetoresistivity $\rho_{xx}$ and Hall resistivity $\rho_{yx}$ in various magnetic fields. (c), (d) Magnetic-field dependences of $\rho_{xx}$ and $\rho_{yx}$ at various temperatures. Inset in (b) shows different contributions to $\sigma_{xy}(T)$ at  $B=10$T. The dashed lines mark the temperatures at which $\mu=0$ and $\sigma_{xy}=0$ respectively.  Units of all numerics are the same as in Fig.~\ref{fig:transport_B0}.  }
\label{fig:rho(TB)}
\end{figure*}

\emph{Resistivity, Hall and Seebeck coefficients.} We now proceed to calculate transport coefficients and see if the above thermodynamic behaviors can result in the experimentally observed resistivity peak and other transport phenomena. We begin with the longitudinal resistivity $\rho_{xx}(T)$, Hall coefficient $R_H(T)$ and Seebeck coefficient $S_{xx}(T)$ in zero magnetic field. Below we only discuss transport properties in electron-doped systems (i.e., $E_F>0$). Zero-field transport in hole-doped systems is discussed in SM\cite{Note1}.

To calculate transport coefficients, we use the Kubo formula (see e.g. Refs.~\onlinecite{MahanBook,RickayzenBook}). Our calculations are standard and details are included in SM\cite{Note1}. The Hall coefficient, defined by $\rho_{yx} = R_H B$, is obtained by taking the $B\rightarrow 0$ limit of our finite-magnetic-field results. The Seebeck coefficient is computed by a generalized Mott formula that was obtained in Ref.~\onlinecite{JonsonPRB1984}.  Here, we discuss how disorder is treated. Disorder is the \emph{only} source of resistivity in our calculations as electron-phonon and electron-electron scattering are absent. It is included by an energy level broadening $\Gamma_a$ in the single-particle Green's function
\begin{equation}
G_a(\omega) = \frac{1}{\omega - E_{a} + i \Gamma_a(\omega)}
\label{eq:green}
\end{equation} 
where $E_a$ is the energy associated with the single-particle eigenstate $|a\rangle$. We take a crude simplification: $\Gamma_a(\omega)=\Gamma_1$ is a constant for all eigenstates of the Dirac fermion, and $\Gamma_a(\omega)=\Gamma_2$ is also a constant for all eigenstates of the quadratic fermions.  We will see that this somewhat oversimplified treatment produces surprisingly good results. In SM\cite{Note1}, we also apply the Born approximation\cite{RickayzenBook} of disorder for zero-field transport quantities as a comparison, but no qualitative difference is observed. Therefore, we will not bother to apply more realistic disorder models at finite $B$ fields where self-consistent Born approximation may be necessary for large $B$\cite{AndoRMP1982}.

Evaluations of $\rho_{xx}(T)$, $R_H(T)$ and $S_{xx}(T)$ are done numerically with an input of  temperature-dependent $\mu$ determined from Eq.~\eqref{eq:density}. The results are shown in Fig.~\ref{fig:transport_B0}. Indeed, an anomalous peak appears in the longitudinal resistivity and the sign reverses in both the Hall and Seebeck coefficients, all of which occur around the temperature $T_0$. The shapes of the curves in Figs.~\ref{fig:transport_B0}(a), \ref{fig:transport_B0}(c) and \ref{fig:transport_B0}(e) agree very well with those in experiments.\cite{AnomalousPeak1,AnomalousPeak2, HallSignReverse1, HallSignReverse2, HallSignReverse3, ChiNJP2017,ZhangNature2019,LiuNatComm2016} To have a better understanding, we show different contributions to the longitudinal conductivity $\sigma_{xx}$ and Hall conductivity $\sigma_{xy}$ in Fig.~\ref{fig:transport_B0}(b) and \ref{fig:transport_B0}(d).  Contributions from intra-branch scatterings of the Dirac fermion ($\sigma_{\alpha\beta,e}^{(1)}$ and $\sigma_{\alpha\beta,h}^{(1)}$) dominate, and those from inter-branch scattering $\sigma_{\alpha\beta,eh}^{(1)}$ and the quadratic fermions $\sigma_{\alpha\beta}^{(2)}$ are negligible in the temperature regime of our interests. Intuitively, the ``relativistic'' Dirac fermion moves much faster than the ``non-relativistic'' quadratic fermions, and so contributes more to the conductivity.\footnote{For a kinetic energy of $15$meV, quadratic fermions have a velocity $\hbar v_{x2}\approx \hbar\sqrt{2\cdot 15 \text{meV} /(3m_x^*)}\approx 0.6$ eV$\cdot$\AA, compared to $\hbar v_{x1} \approx 6$ eV$\cdot$\AA \ for the Dirac fermion.}  Although they do not conduct much current, the quadratic fermions do serve as good thermodynamic reservoirs, as shown in Fig.~\ref{fig:transport_B0}(f). That is, they are thermodynamically activated, but not quite in transport.  

Analytically, once we neglect the contributions from the quadratic fermions and inter-branch scattering of the Dirac fermion and focus on the regime $ T\gg \Gamma_1$, the conductivities can be approximated by 
\begin{align}
\sigma_{xx}& \approx  C_{xx} \sum_{\lambda=\pm}\int_m^\infty \!\!\! d\epsilon \  \frac{(\epsilon^2-m^2)^{3/2}}{\epsilon} [-f_T'(\lambda\epsilon-\mu)],\nonumber\\
\sigma_{xy} & \approx C_{xy}\sum_{\lambda=\pm}\int_m^\infty \!\!\! \!d\epsilon \ \frac{(\epsilon^2-m^2)^{3/2}}{\epsilon^2} [\lambda f_T'(\lambda\epsilon-\mu)],
\end{align}
where the coefficients are $C_{xx}=v_x^2 \alpha_1\hbar e^2/(3\Gamma_1)$ and $C_{xy}=e^3\hbar^2 B v_x^2 v_y^2 \alpha_1 / (6c\Gamma_1^2)$. The conductivity $\sigma_{yy}$ can be obtained by replacing the index ``$x$'' with ``$y$'' in the expression of $\sigma_{xx}$, and the resistivity is given by $\rho_{\alpha\beta}=(\sigma^{-1})_{\alpha\beta}$.  One can see that when $\mu=0$, the Hall conductivity $\sigma_{xy}$ is zero, which is a consequence of the PH symmetry of the Dirac fermion. The non-zero conductivity $\sigma_{xy}^{(2)}$, though tiny, makes the sign-reversing temperature $\tilde{T}_0$ of $R_H$ differ slightly away from the sign-reversing temperature $T_0$ of $\mu$.

\emph{Magnetoresistivity and Hall resistivity at finite $B$.} The longitudinal resistivity $\rho_{xx}(T,B)$ and Hall resistivity $\rho_{yx}(T,B)$ in a finite magnetic field $B$ are also calculated. The calculation is similar to the zero-field case: We first express the Kubo formula of the conductivity tensor $\sigma_{\alpha\beta}$ in the Landau level basis, then input the chemical potential $\mu(T,B)$ obtained from Eq.~\eqref{eq:density}, and finally evaluate the conductivity numerically (see details in SM \cite{Note1}). 

Numerical results are shown in Fig.~\ref{fig:rho(TB)}. Our focus is the high-field and high-temperature regime, i.e., $T, \hbar\omega_{c1}\gtrsim \Gamma_1$.  The curves of $\rho_{xx}$ and $\rho_{yx}$ as functions of $T$ or $B$ again show good agreement with experiments. Two features deserve some attention. First, the ``anomalous'' peak in $\rho_{xx}(T)$ is largely enhanced by the magnetic field, which was initially observed in experiments in  Ref.~\cite{TrittPRB1999}(see also \cite{LiuNatComm2016,ZhangNature2019}). Theoretically, current conduction occurs  when electrons/holes hop between two states, in the $N$-th and $(N+1)$-th Landau levels respectively, whose energies overlap after disorder broadening. By increasing $B$ such that Landau level spacing is larger than $\Gamma_1$, available states that overlap in energy greatly decrease, leading to enhancement of resistivity. For the same reason, $\rho_{yx}$ is also enlarged by the magnetic field. Second, the temperature $\tilde T_0$ at which  $\sigma_{xy}=0$ (equivalently ${\rho_{yx}=0}$) increases as $B$ increases (e.g., see experiments in Refs.~\cite{LiuNatComm2016,ZhangNature2019}). When $B$ is large, $\sigma_{xy}^{(2)}$ cannot be neglected as shown in the inset of Fig.~\ref{fig:rho(TB)}(b). The underlying reason is that Landau level spacing is much smaller for the quadratic fermions than for the Dirac fermion, so $\sigma_{xy}^{(1)}$ reduces faster than $\sigma_{xy}^{(2)}$ as $B$ increases.  If $\sigma_{xy}^{(2)}$ is neglected, $\sigma_{xy}=0$ occurs at $\mu=0$ and so $\tilde T_0= T_0$ which has negligible $B$ dependence. Now that $\sigma_{xy}^{(2)}$ is non-negligible, its  $B$ dependence as well as the $B$ dependence of other conductivity contributions  make $\tilde{T}_0$ increase as $B$ increases. This is also the reason behind the feature that $\rho_{yx}(B)$ reverses the sign as $B$ increases, for certain temperatures, as shown in Fig.~\ref{fig:rho(TB)}(d).

\emph{Discussions.} In summary, we have proposed a mechanism for the long-standing transport anomaly in dilute metals ZrTe$_5$ and HfTe$_5$. Our proposal describes a scenario that a minority current carrier may be thermodynamically very active, leading to intriguing interplay between equilibrium thermodynamics and transport properties. Many aspects of this work can be improved, e.g., by a more realistic handling of disorder and by including electron-phonon coupling in high-temperature regime. However, we believe that our model captures the essence of the experimentally observed transport anomalies. In the present model, the quadratic bands are responsible for PH symmetry breaking. However, in real samples there exist other factors that break PH symmetry, e.g., additional electron pockets\cite{NarayanPRR2019} or Dirac band itself being asymmetric\cite{FuPRL2020}. Hence, detailed experimental or first-principles study on the asymmetry of low-energy band structure is strongly encouraged, for the purpose of verifying or falsifying our theory at a more quantitative level. This work can be thought of as a microscopic theory of the phenomenological multi-carrier model that is commonly used to fit experimental data.\cite{ChiNJP2017,LiuNatComm2016,ZhangNature2019} For future studies, it would be interesting to extend this work to other dilute metals such as SrTiO$_3$\cite{StemmerReview2018}.

\emph{Acknowledgments.} CW acknowledges Liyuan Zhang for introducing the problem studied in this work and for various enlightening discussions and a careful reading of the manuscript, without whom this work would never have occurred. CW is grateful to Shizhong Zhang for encouragement and helpful discussions, and to Shun-Qing Shen for valuable discussions and for sharing his manuscript before publication.  This work was supported by Research Grant Council of Hong Kong (ECS 21301018 and GRF 11300819) and URC, HKU (Grant No. 201906159002).

\bibliography{qhe.bib}


\renewcommand{\thefigure}{S\arabic{figure}}
\setcounter{secnumdepth}{3}
\setcounter{figure}{0}


\onecolumngrid

\appendix

\begin{figure}[b]
\centering
\includegraphics[scale=1]{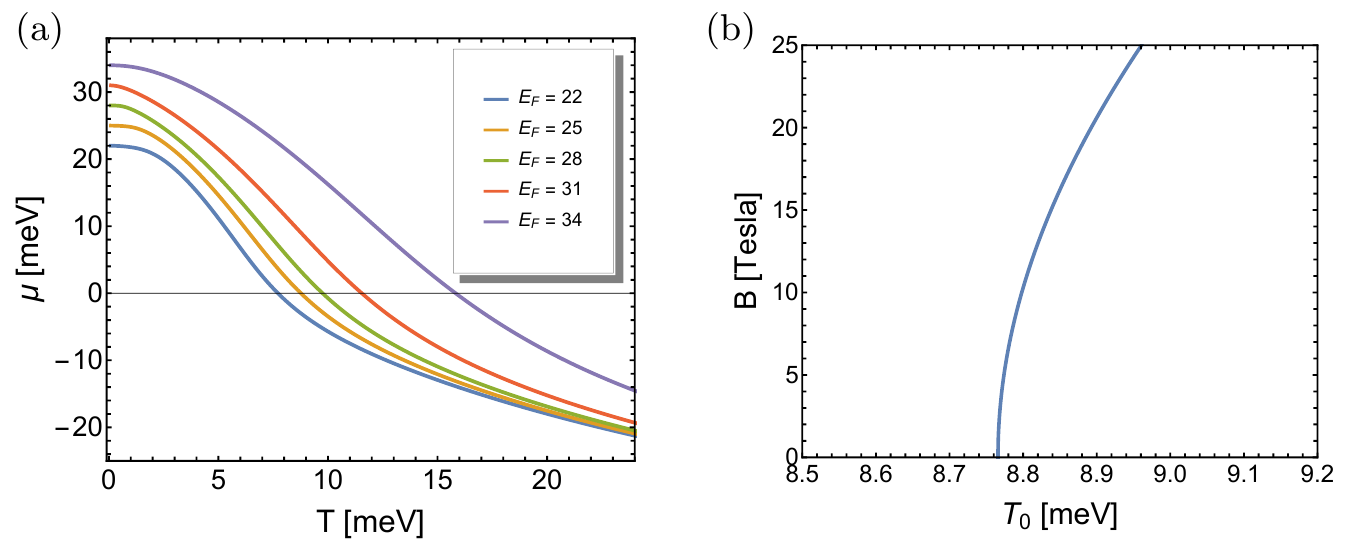}
\caption{ (a) Zero-field chemical potential $\mu(T)$ with different Fermi energies $E_F$ in our model. Note that $E_F=31$ and $34$ meV are above the band bottom $\Delta=30$ meV of the quadratic fermions. (b) The temperature $T_0(B)$ with $E_F=25$ meV, at which $\mu$ reverses its sign. All other parameters are set to the same numerical values as in Fig.~\ref{fig:muT}, if not mentioned here.}
\label{fig:Tc}
\end{figure}

\section{Properties of $\mu(T,B)$}
In this section, we discuss some technical details and show additional properties regarding the chemical potential $\mu(T,B)$. 

\subsection{Monotonicity of $\mu(T)$ at $B=0$}
The chemical potential $\mu(T,B)$ is defined through the integral equation \eqref{eq:density} by imposing a constant density $n$. Here we prove a general monotonic behavior of $\mu(T)\equiv \mu(T,B=0)$. We claim that $\mu(T)$ is monotonically decreasing, if $\mu(T)$ itself is non-negative and if the DOS $D(\epsilon)$ satisfies the following two conditions: (i) $D(\epsilon)$ is monotonically increasing for $\epsilon\ge 0$, and monotonically decreasing for $\epsilon\le 0$ and (ii) $D(|\epsilon|)-D(-|\epsilon|)\ge 0$. The two conditions are satisfied in our low-energy model of ZrTe$_5$. 

To prove the claim, we need to show $\partial \mu/\partial T |_{n}\le 0$ when $\mu \ge 0$. This derivative can be alternatively written as
\begin{equation}
\left.\frac{\partial \mu}{\partial T}\right|_{n} = -\left.\frac{\partial n}{\partial T}\right|_{\mu} \cdot \left( \left.\frac{\partial n}{\partial \mu}\right|_{T} \right)^{-1}
\label{eq:app1}
\end{equation}
With the expression \eqref{eq:density} of the density $n(T,\mu)$, derivatives on the right-hand side of \eqref{eq:app1} can be easily obtained. We have
\begin{align}
\left.\frac{\partial n}{\partial T}\right|_{\mu} & = -\int_{-\infty}^\infty d\epsilon D(\epsilon) f_T'(\epsilon-\mu) \frac{\epsilon-\mu}{T}= \int_0^\infty d\epsilon [D(\epsilon+\mu) - D(-\epsilon+\mu)][-f_T'(\epsilon)] \frac{\epsilon}{T} 
\end{align}
where $f_T'(x) = \partial f_T(x)/\partial x$. It is not hard to see that the integrand is always non-negative for $\epsilon\ge 0$, provided that conditions (i) and (ii) are satisfied. In addition, 
\begin{align}
\left.\frac{\partial n}{\partial \mu}\right|_{T} & = \int_{-\infty}^\infty d\epsilon D(\epsilon)[-f_T'(\epsilon-\mu)]>0
\end{align}
Combining all together, we immediately have $\partial \mu/\partial T|_n \le  0$. The equality is achieved only in special cases, e.g., $D(\epsilon)$ is a constant function.

With this monotonic property, we see that starting at a positive $E_F$, $\mu$ will monotonically decrease as $T$ increases, until it becomes negative (it may continue to decrease after being negative). One may prove similar results  in the opposite scenario of a negative $E_F$. We comment that this monotonicity extends to a finite $B$ field roughly when $T\gtrsim \hbar\omega_c$, where $\hbar\omega_c$ denotes a characteristic cyclotron energy. When $T\lesssim \hbar \omega_c$, the monotonicity is affected by quantum oscillation in $\mu$.

\subsection{Sign-reversing temperature $T_0$}

The sign-reversing temperature $T_0$, defined by $\mu(T_0,B)=0$, depends on $n$, $B$ and a few other quantities. In experiments, different samples have different densities and so have different $T_0$'s . Fig.~\ref{fig:Tc}(a) shows $\mu(T)$ at $B=0$ for different $n$'s. We observe that $T_0$ increases as $n$ increases (equivalently, as $E_F$ increases). 

The magnetic field dependence $T_0(B)$ is quadratic to the lowest order [Fig.~\ref{fig:Tc}(b)]. To see that analytically, let us obtain an expression of $T_0(B)$ for a small $B$ field.  Since $D_1(\epsilon,B)$ is PH symmetric, at $\mu=0$, Eq.~\eqref{eq:density} reduces to
\begin{equation}
n= \int_0^\infty d\epsilon D_2(\epsilon,B)f_T(\epsilon) = \kappa\alpha_2 \sum_{N\ge 0} \int_0^\infty d\epsilon  \frac{\hbar \omega_{c2}}{\sqrt{\epsilon}} f_T(\epsilon + E_{N2})
\label{eq:app2}
\end{equation}
This equation determines $T_0(B)$. Recall that $\hbar\omega_{c2}=eB/c\sqrt{m_x^*m_y^*}$ and $E_{N2} = \hbar \omega_{c2}(N+1/2) + \Delta$. In the limit $B\rightarrow 0$, the summation over $N$ can be replaced by an integral
\begin{equation}
n=\kappa \alpha_2 \int_0^\infty d\epsilon d\omega  \frac{1}{\sqrt{\epsilon}} f_T(\epsilon + \omega+\Delta)
\label{eq:app3}
\end{equation}
which determines the $T_0\equiv T_0(0)$. Accordingly,  $\delta T_0=T_0(B)-T_0$ is the difference between the summation \eqref{eq:app2} and the integral  \eqref{eq:app3}. For a small $B$, this can be evaluated by Taylor expanding $f_T(\epsilon+E_{N2})$ in \eqref{eq:app2} around $B=0$ and $T=T_0$, and then take the continuum limit. One can check that  the lowest order non-vanishing term is of $B^2$. After some calculations, we obtain 
\begin{align}
\delta T_0 =  \frac{\hbar^2\omega_{c_2}^2}{24T_0} g(\Delta/T_0)
\end{align}
where $g$ is a dimensionless function
\begin{equation}
g(z) = \frac{\int_0^\infty dx f'(x+z)/\sqrt{x}}{ \int_0^\infty dx d y [(x+y+z)/\sqrt{x}]f'(x+ y+z) }  
\end{equation}
Here $f(x) = 1/(e^x + 1)$ and $f'(x)$ is its first derivative. The function $g(z)$ is positive valued and monotonically decreasing as $z$ increases. For numerical values in the caption of Fig.~\ref{fig:muT}, $\Delta/T_0 \approx 3.4$ and then $\delta T_0 \approx 3.3\times 10^{-4} B^2$ with $B$ in units of Tesla and $\delta T_0$ in meV.

We comment that if the Landau level energy is $\hbar \omega_{c2}(N+\nu) + \Delta$ with $\nu\neq 1/2$, the lowest order term in $\delta T_0$ will be linear in $B$. More precisely, $\delta T_0\propto (\nu-1/2)B$. Accordingly, $\nu=1/2$ makes the linear correction vanishing. The situation $\nu\neq 1/2$ occurs in the presence of non-trivial Berry phase in Bloch wave functions.

\section{Methods for computing transport coefficients}
In this section, we describe details of the calculations on the transport coefficients. Most of our calculations are standard applications of the celebrated Kubo formula.

\subsection{Generalities}

We use the standard Kubo formula to compute the dc conductivity tensor $\sigma_{\alpha\beta}$, with $\alpha,\beta=x,y$ (see, for example, Ref.~\onlinecite{MahanBook}). Conductivities, both at zero and finite magnetic fields, will be calculated. We will neglect inter-band scattering between the Dirac and quadratic fermions, as well as scattering between the four copies of quadratic fermions, however contribution from scattering between the particle and hole branches of the Dirac fermion will be included (though it turns out to be very small compared to that from intra-branch scattering). The total conductivity is the sum of $\sigma_{\alpha\beta}^{(1)}$ and $\sigma_{\alpha\beta}^{(2)}$, of the Dirac and quadratic fermions respectively. For the current general discussion, we omit the superscript of $\sigma_{\alpha\beta}^{(i)}$ for clarity. According the Kubo formula, the conductivity tensor is given by
\begin{equation}
\sigma_{\alpha\beta} = \lim_{\omega \rightarrow 0} \frac{\hbar}{\omega} \mathrm{Im} \left[ K_{\alpha\beta}(\omega+i\delta)- K_{\alpha\beta}(0)\right] 
\label{eq:sigma-general}
\end{equation}
where $K_{\alpha\beta}$ is the retarded Green's function of the current operator $J_{\alpha}=-e\hat v_\alpha$, with $\hat v_\alpha$ being the velocity operator and $-e$ being the electron charge. The velocity operator is given by  $\hat v_\alpha= i[H,x_\alpha]/\hbar$ , where $H$ is the Hamiltonian and $x_\alpha$ is the position operator. $K_{\alpha\beta}(0)$ represents the diamagnetic response.  In terms of Matsubara frequency $i\omega_n = i2\pi nT$, we have $K_{\alpha\beta}(i\omega_n)= \frac{1}{V} \int_0^{1/T} \langle J_\alpha(\tau) J_\beta(0)\rangle $, where $V$ is the volume. The retarded Green's function is then obtained by analytic continuation $i\omega_n\rightarrow \omega + i\delta$. Writing the Green's function in the spectral representation, one finds that
\begin{align}
K_{\alpha\beta}(i\omega_n) = -\frac{e^2\hbar}{V}\sum_{ab} (v_\alpha)_{ab} (v_\beta)_{ba} \int\frac{d\epsilon_1}{2\pi}\frac{d\epsilon_2}{2\pi} \mathcal{A}_a(\epsilon_1) \mathcal{A}_b(\epsilon_2) \frac{f_T(\epsilon_1)-f_T(\epsilon_2)}{i\omega_n+\epsilon_1-\epsilon_2}
\label{eq:Green}
\end{align}
where $(v_\alpha)_{ab} = \langle a|\hat{v}_\alpha|b\rangle$ is the matrix element between two single-particle eigenstates $\{|a\rangle\}$, $\mathcal{A}_a(\epsilon)$ is the spectral function, and $f_T(\epsilon)$ is the Fermi-Dirac distribution.  In the expression \eqref{eq:Green}, vertex correction to the two-particle Green's function is neglected (which indeed vanishes in the case of Born approximation for $\delta$-function impurities\cite{AndoRMP1982}).  

After analytic continuation $i\omega_n \rightarrow \omega +i\delta$, inserting \eqref{eq:Green} into \eqref{eq:sigma-general}, taking the limit $\omega\rightarrow 0$, one can show that the conductivity tensor is given by
\begin{align}
\sigma_{\alpha\beta} = \frac{e^2 \hbar }{ V} \sum_{ab} \left\{\mathrm{Re}[(v_\alpha)_{ab} (v_\beta)_{ba}] \mathcal{P}_{ab} - \mathrm{Im}[(v_\alpha)_{ab} (v_\beta)_{ba}] \mathcal{Q}_{ab}\right\}
\label{eq:conductivity-formula}
\end{align}
where we have defined
\begin{align}
\mathcal{P}_{ab}  =  \int \frac{d\omega}{4\pi} \mathcal{A}_a(\omega)\mathcal{A}_b(\omega)[-f'_T(\omega)], \quad 
\mathcal{Q}_{ab}  = \int \frac{d\omega}{2\pi} [\mathcal{A}_a(\omega) \mathcal{R}_b'(\omega) -\mathcal{A}_b(\omega) \mathcal{R}_a'(\omega)] f_T(\omega)
\label{eq:PQ}
\end{align}
where $\mathcal{R}_a(\omega)$ is the real part of the single-particle Green's function, and $\mathcal{R}'_a$ is the first derivative of $\mathcal{R}_a(\omega)$. This formula can also be found in Ref.~\cite{JonsonPRB1984}. Note that $\mathcal{P}_{ab}=\mathcal{P}_{ba}$ and $\mathcal{Q}_{ab} = -\mathcal{Q}_{ba}$.  

We will assume that the single-particle Green's function has the following form
\begin{equation}
G_a(\omega)= \frac{1}{\omega-E_a + i \Gamma}
\end{equation}
for a single-particle eigenstate $|a\rangle$ with eigenenergy $E_a$. We take a crude assumption that the imaginary part $\Gamma$ of the self-energy is a constant ($\Gamma_1$ and $\Gamma_2$ for the Dirac and quadratic fermions respectively). Then,
\begin{equation}
\mathcal{A}_a(\omega) =\mathcal{A}(E_a-\omega) = \frac{2\Gamma}{(\omega-E_a)^2 + \Gamma^2}, \quad \mathcal{R}_a(\omega)= \mathcal{R}(E_a-\omega) = \frac{\omega - E_a}{(\omega-E_a)^2 + \Gamma^2}
\end{equation}
This means $\mathcal{P}_{ab}=\mathcal{P}(E_a,E_b)$ and $\mathcal{Q}_{ab}=\mathcal{Q}(E_a,E_b)$. The level broadening $\Gamma$ is due to disorder scattering. For weak disorder, one generally uses Born approximation to obtain the self-energy, which leads to an energy dependent $ \Gamma(\omega)$ that is proportional to the density of states $ D(\omega)$ per band. Moreover, in the presence of strong magnetic field, self-consistent Born approximation is preferred\cite{AndoRMP1982}. Nevertheless, even with a constant $\Gamma$, we find our results can reproduce the main features of resistivity and magnetotransport observed in experiments.  So, we will stick to constant $\Gamma$ for simplicity. (At the same time, we  do have computed the resistivity and Hall coefficient at zero field using Born approximation for comparison, which are discussed in Appendix \ref{sec:app-Born}.) We estimate the ratio $\Gamma_2/\Gamma_1$ by taking the Born approximation which gives
\begin{align}
\frac{\Gamma_2}{\Gamma_1} = \frac{D_2(E_2)/\kappa}{D_1(E_1)}
\end{align}
for properly chosen $E_2$ and $E_1$, and $D_1(\epsilon), D_2(\epsilon)$ are given in Eq.~\eqref{eq:dos}. We chose $E_1=E_F$ and $E_2$ to be the average energy of the quadratic fermions at $T=T_0$. With the numerics in Fig.~\ref{fig:muT}, we have $\Gamma_2/\Gamma_1\approx 7$. In our calculations, we set $\Gamma_2/\Gamma_1 = 10$. The specific ratio does not affect the results qualitatively on transport properties as long as it is not too small.

In the limit $\Gamma\ll T $, the function $f_T(\omega)$ is a smooth function compared with $\mathcal{A}$ and $\mathcal{R}$. Then, one can show that the following approximations hold
\begin{align}
\mathcal{P}_{ab} \approx \left\{
\begin{array}{ll}
 \mathcal{A}^{\Gamma}_b(E_a)[-f_T'(E_a)-f_T'(E_b)]/2, & E_a-E_b \gg \Gamma\\[5pt]
 \mathcal{A}^{2\Gamma}_b(E_a)[-f_T'(E_a)]/2, &  E_a-E_b \ll \Gamma
\end{array}
\right.
\label{eq:p-simp}
\end{align}
where $\mathcal{A}_b^{2\Gamma}$ means the spectral function with $\Gamma$ replaced by $2\Gamma$. Also, 
\begin{align}
\mathcal{Q}_{ab} \approx \left\{
\begin{array}{ll}
 \mathcal{R}'_b(E_a)f_T(E_a)-\mathcal{R}'_a(E_b)f_T(E_b), & E_a-E_b \gg \Gamma\\[5pt]
 \frac{(E_a-E_b)^3 + 4\Gamma^2 (E_a-E_b)}{[(E_a-E_b)^2 + 4\Gamma^2]^2} [-f_T'(E_b)], &  E_a-E_b \ll \Gamma
\end{array}
\right.
\label{eq:q-simp}
\end{align}
Note that the energies $E_a,E_b$ in the above expressions are measured with respect to the chemical potential $\mu$.

\subsection{Transport at $B=0$}
\label{sec:app_transportB=0}
Let us now compute transport coefficients in zero field. We compute the temperature dependence of the resistivity $\rho_{xx}$, Hall coefficient $R_H$, and Seebeck coefficient $S_{xx}$.  The Dirac fermion is described by the Hamiltonian 
\begin{align}
H =  m \tau_z \sigma_0 + v_x \hat p_x \tau_x \sigma_z + v_y \hat p_y \tau_y \sigma_0 + v_z \hat p_z \tau_x \sigma_x
\label{eq:DiracH}
\end{align}
where $\sigma_i$ and $\tau_i$ are Pauli matrices with $\sigma_i$ acting on spins. It is straightforward to obtain the eigenstates $|\bfk\lambda\sigma\rangle$, where $\lambda=\pm1$ denotes the particle and hole branches respectively and $\sigma=\pm1$ denotes the spin state. The corresponding energy is $E_{\bfk\lambda\sigma} = \lambda \sqrt{\hbar^2v_x^2 k_x^2+ \hbar^2v_y^2 k_y^2+ \hbar^2v_z^2 k_z^2 + m^2 }$, which is independent of the spin index $\sigma$ since we have neglected the Zeeman coupling. Eigenstates of the anisotropic quadratic fermion are denoted as $|\bfk\sigma\rangle$ and the energy is $E_{\bfk\sigma} = \hbar^2k_x^2/2m_x^* + \hbar^2k_y^2/2m_y^* + \hbar^2k_z^2/2m_z^* + \Delta$, where $\Delta$ is the bottom of the dispersion. According to Eq.~\eqref{eq:conductivity-formula}, the longitudinal conductivity $\sigma_{xx}$ can then be expressed as
\begin{equation}
\sigma_{xx} = \hbar e^2 \sum_{\nu\nu'}\int \frac{d^3\mathbf{k}}{(2\pi)^3} \left|\langle \mathbf{k}\nu|\hat{v}_x|\mathbf{k}  \nu'\rangle\right|^2 \mathcal{P}(E_{\bfk\nu}-\mu, E_{\bfk\nu'}-\mu)
\label{eq:appsxxB0}
\end{equation}
where $\nu=\lambda\sigma$ or $\nu=\sigma$ for the Dirac and quadratic fermions respectively. In this expression, we have used the fact that $\hat{v}_x = i[H,x]/\hbar$ is diagonal in the index $\bfk$, and have  written out the the chemical potential $\mu$ explicitly.  Calculations of the matrix element $\langle \mathbf{k}\nu|\hat{v}_x|\mathbf{k}  \nu'\rangle$ is straightforward. After some simplifications, we obtain
\begin{align}
\sigma_{xx}^{(1)} & = \hbar e^2 v_x^2 \sum_{\lambda\lambda'} \int_0^\infty d\epsilon D_1(\epsilon) \mathcal{M}_{1,\lambda\lambda'}(\epsilon )  \mathcal{P}(\lambda\epsilon-\mu, \lambda'\epsilon-\mu)
\label{eq:sigmaxx1}\\
\sigma_{xx}^{(2)} & = \frac{\hbar e^2}{m_x^*} \int_0^\infty d\epsilon D_2(\epsilon) \mathcal{M}_{2}(\epsilon ) \mathcal{P}(\epsilon-\mu,\epsilon-\mu)
\label{eq:sigmaxx2}
\end{align}
where $\sigma_{xx}^{(1)}$ and $\sigma_{xx}^{(2)}$ are contributions from the Dirac and quadratic fermions respectively,  $D_i(\epsilon)$ are DOS's given in Eq.~\eqref{eq:dos},  and 
\begin{equation}
\mathcal{M}_{1,\lambda\lambda'}(\epsilon) = \frac{1}{2} -\frac{\lambda\lambda'}{2}\frac{\epsilon^2 + 2m^2}{3\epsilon^2}, \quad \mathcal{M}_{2}(\epsilon )  = \frac{2}{3}(\epsilon-\Delta)
\end{equation}
The total conductivity is given by $\sigma_{xx} = \sigma_{xx}^{(1)} +\sigma_{xx}^{(2)}$. The conductivities $\sigma_{yy}^{(i)}$ can be obtained simply by replacing $x$ with $y$ in the above expressions.

The Hall conductivity $\sigma_{xy}$ under a finite magnetic field $B$ is obtained below in Eqs.~\eqref{eq:sigma1xyfiniteB} and \eqref{eq:sigma2xyfiniteB} in Sec.~\ref{sec:appfiniteB}. We derive the Hall coefficient simply by considering small $B$ and separating out the piece that is linear in $B$. We obtain the following expressions
\begin{align}
\sigma_{xy}^{(1)} & = \frac{e^3\hbar^2B}{c} v_x^2v_y^2 \sum_{\lambda\lambda'} \int_0^\infty d\epsilon D_1(\epsilon)  \frac{\lambda' \mathcal{M}_{1,\lambda\lambda'}(\epsilon )}{\epsilon}  \frac{\partial\mathcal{Q}}{\partial E_b}(\lambda\epsilon-\mu, \lambda'\epsilon-\mu)
\label{eq:sigmaxy1}\\
\sigma_{xy}^{(2)} & = \frac{e^3\hbar^2B}{c} \frac{1}{m_x^*m_y^*} \int_0^\infty d\epsilon D_2(\epsilon) \mathcal{M}_{2}(\epsilon ) \frac{\partial\mathcal{Q}}{\partial E_b}(\epsilon-\mu, \epsilon-\mu)
\label{eq:sigmaxy2}
\end{align}
where $\partial \mathcal{Q}(E_a,E_b)/\partial E_b$ is the first derivative of $\mathcal{Q}(E_a,E_b)\equiv \mathcal{Q}_{ab}$ defined in \eqref{eq:PQ}. The total Hall conductivity is $\sigma_{xy} = \sigma_{xy}^{(1)} + \sigma_{xy}^{(2)}$, and $\sigma_{yx}=-\sigma_{xy}$. The resistivity tensor is given by $\rho_{\alpha\beta} = (\hat \sigma^{-1})_{\alpha\beta}$ and the Hall coefficient $R_H$ is given by $\rho_{yx}= R_H B$. Numerical evaluations of the expressions \eqref{eq:sigmaxx1}, \eqref{eq:sigmaxx2}, \eqref{eq:sigmaxy1} and \eqref{eq:sigmaxy2} are performed with varying temperature $T$ and the chemical potential $\mu(T)$ determined by Eq.~\eqref{eq:density}. The results are shown in Fig.~\ref{fig:transport_B0}. 

The above expressions of $\sigma_{xx}^{(1)}$ and  $\sigma_{xy}^{(1)}$ can be simplified using the following facts: (i) We will set the level broadening $\Gamma_1=0.5$ meV, which satisfies $\Gamma_1\ll T$ in most part of the temperature regime that we are interested in,  so the approximations \eqref{eq:p-simp} and \eqref{eq:q-simp} are applicable;  (ii) the inter-branch contributions to $\sigma_{xx}^{(1)}$ is small compared to intra-branch contributions in the whole temperature regime that we are interested in, and so they can be neglected. The simplification has been checked numerically and works very well. Then, we have the following simplified expressions 
\begin{align}
\sigma_{xx}^{(1)} & \approx \frac{v_x^2 \alpha_1\hbar e^2}{3\Gamma_1} \int_m^\infty d\epsilon \frac{(\epsilon^2-m^2)^{3/2}}{\epsilon} [-f_T'(\epsilon-\mu)-f_T'(-\epsilon-\mu)]\nonumber\\
\sigma_{xy}^{(1)} & \approx -\frac{e^3\hbar^2 B}{c} \frac{v_x^2 v_y^2\alpha_1}{6\Gamma_1^2}\int_m^\infty d\epsilon \frac{(\epsilon^2-m^2)^{3/2}}{\epsilon^2} [-f_T'(\epsilon-\mu) + f_T'(-\epsilon-\mu)]
\label{eq:sigma-simplified}
\end{align}
Similar expressions can also be obtained for $\sigma_{xx}^{(1)}$ and $\sigma_{xx}^{(2)}$. However, we find that $\sigma_{xx}^{(2)}\ll \sigma_{xx}^{(1)}$ and $\sigma_{xy}^{(2)}\ll \sigma_{xy}{(1)}$ in the temperature of our interests.  Then, conductivities in \eqref{eq:sigma-simplified} dominate. In particular, it is easy to see that $\sigma_{xy}^{(1)}=0$ when $\mu =0$ (which follows the particle-hole symmetry of the Dirac fermion), which indicates a sign reverse of $R_H$ when $\mu$ crosses zero.

To compute the Seebeck coefficients $S_{xx}$, we make use of the generalized Mott formula obtained in Ref.~\cite{JonsonPRB1984} for non-interaction fermion: $\hat S = \hat{\rho} \hat{\varepsilon}$, where
\begin{align}
\varepsilon_{ij} = -\frac{1}{eT} \int_{-\infty}^\infty d\epsilon (\epsilon-\mu) [-f_T'(\epsilon-\mu)]\sigma_{ij}(0,\epsilon)
\end{align}
where $\sigma_{ij}(0,\epsilon)$ is the conductivity at $T=0$ and chemical potential $\mu=\epsilon$. We will only compute the zero-field Seebeck coefficient $S_{xx}=\rho_{xx}\varepsilon_{xx}$. With the expressions of $\sigma_{xx}^{(1)}$ and $\sigma_{xx}^{(2)}$, the conductivity $\sigma_{xx}(0,\epsilon)$ can be easily evaluated, from which we obtain
\begin{align}
\varepsilon_{xx} \approx & -\frac{v_x^2 \hbar e\alpha_1}{3T\Gamma_1} \int_m^\infty d\epsilon  \frac{(\epsilon^2-m^2)^{3/2}}{\epsilon} [-(\epsilon-\mu)f_T'(\epsilon-\mu)+(\epsilon+\mu)f_T'(-\epsilon-\mu)]  \nonumber\\
&  \quad -\frac{2\hbar e\kappa\alpha_2}{3T\Gamma_2 m_x^*} \int_\Delta^\infty d\epsilon (\epsilon-\Delta)^{3/2} [-(\epsilon-\mu)f_T'(\epsilon-\mu)]
\end{align}
In the second line which originates from $\sigma_{xx}^{(2)}$, we have assumed $\Gamma_2\ll T$, which is not quite true. However, $\sigma_{xx}^{(2)}$ is much smaller than $\sigma_{xx}^{(1)}$, so this approximation does not matter too much. If the second line is neglected, we see that $\varepsilon_{xx} = 0$ when $\mu=0$. Accordingly, a sign reverse is expected when $\mu$ crosses zero, similarly to $R_H$.

\subsection{Transport with finite $B$}
\label{sec:appfiniteB}

\begin{figure}[b]
\centering
\begin{tikzpicture}
\node at (0,0){\includegraphics[scale=0.5]{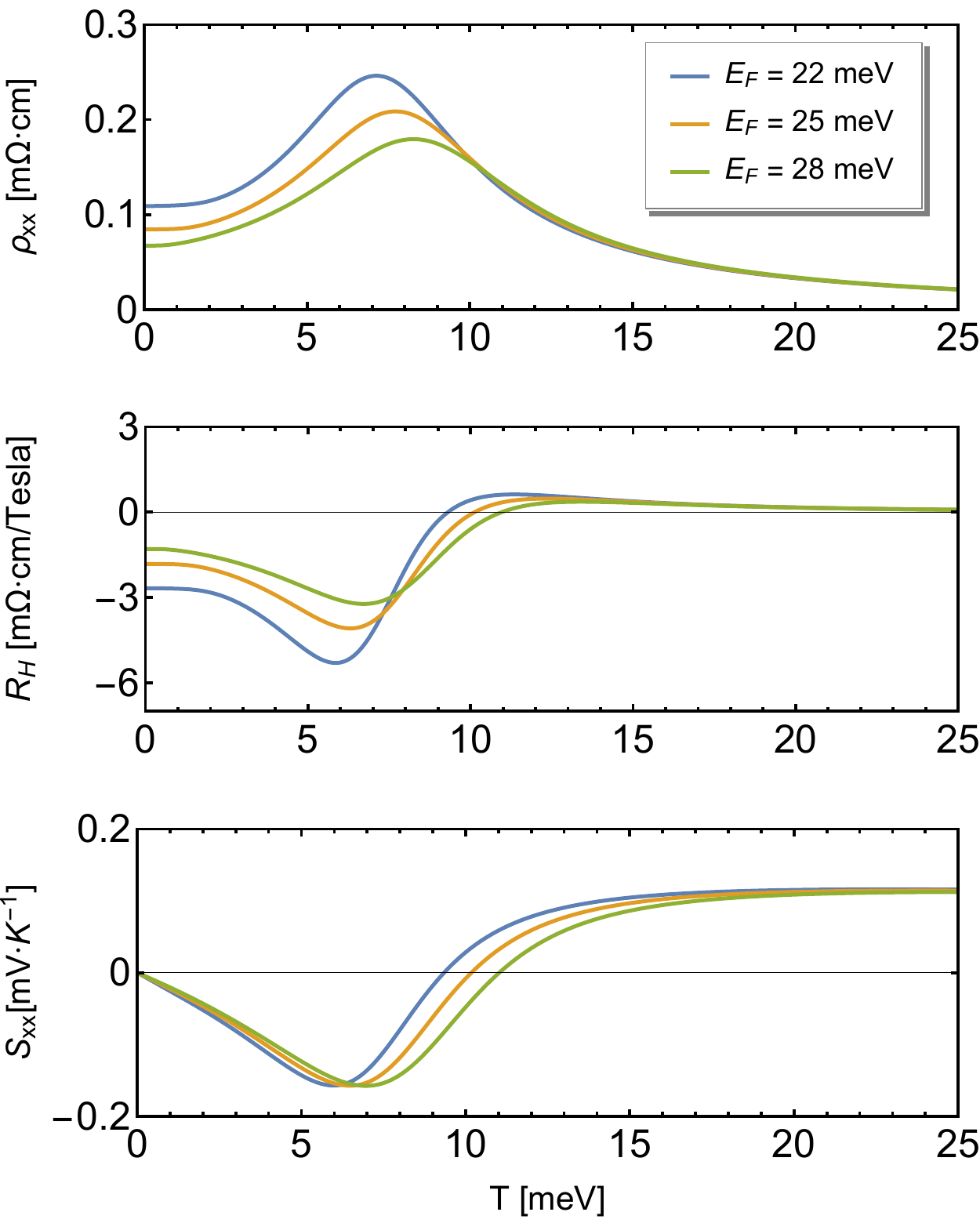}};
\node at (-3,4.2)[scale=1]{\bf (a)};
\node at (8,0){\includegraphics[scale=0.5]{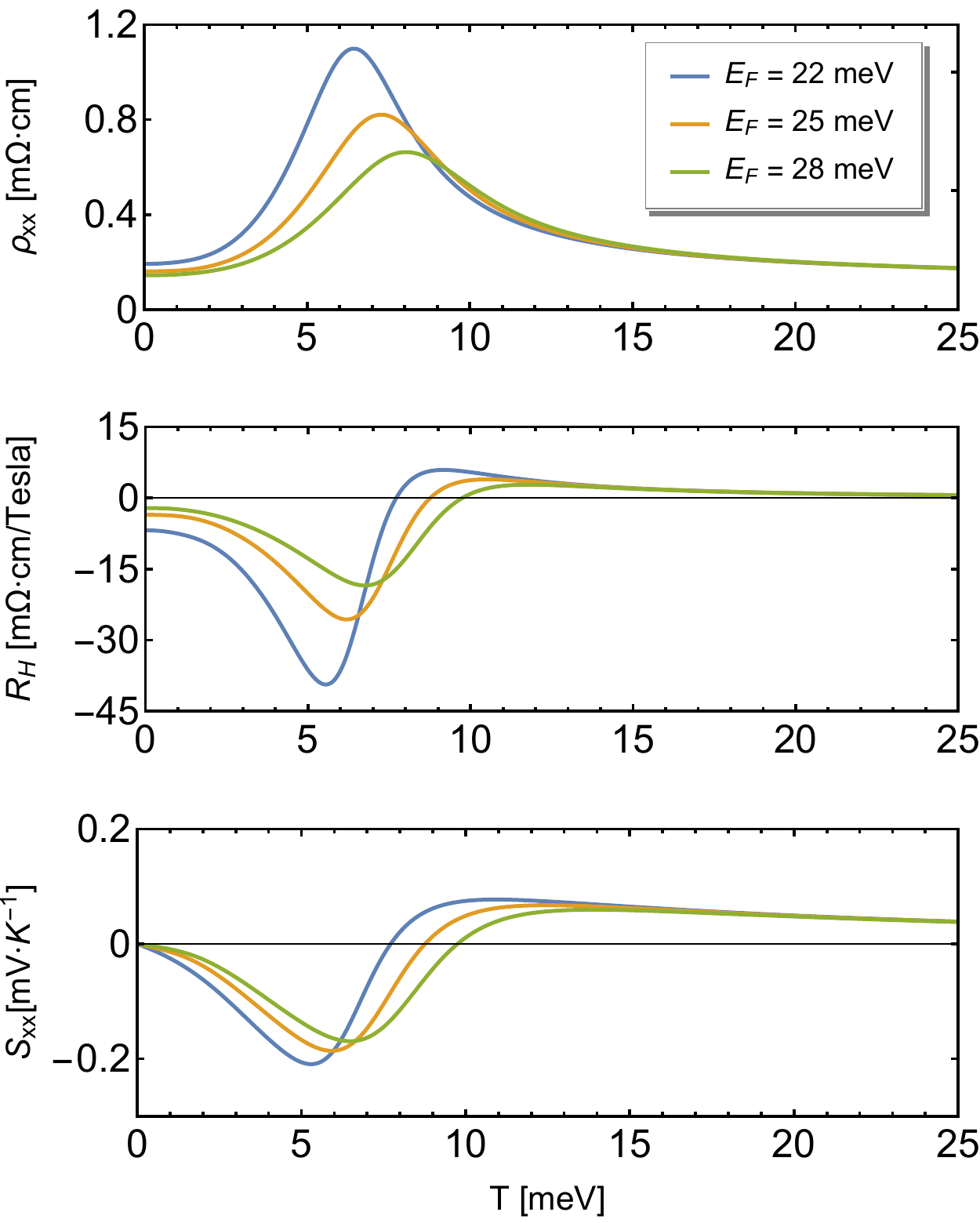}};
\node at (5,4.2)[scale=1]{\bf (b)};
\end{tikzpicture}
\caption{(a) Temperature dependences of $\rho_{xx}$, $R_H$ and $S_{xx}$ for massless Dirac fermion. Numerical data is the same as in Fig.~\ref{fig:transport_B0} except $m=0$ here. (b) Temperature dependences of $\rho_{xx}$, $R_H$ and $S_{xx}$ for Born approximation of disorder.}
\label{fig:additional}
\end{figure}

Now we consider magnetotransport in a uniform magnetic field $B$ along the $\hat{z}$ direction. The magnetic field is included by the minimal coupling $\mathbf{p} \rightarrow \bfp + e\mathbf{A}/c$ in the Hamiltonians. We work in the Landau gauge $\mathbf{A}  = (0, Bx, 0)$. The eigenstates (Landau level states) are well known for both Dirac and quadratic fermions, and can be denoted by $|Nk_yk_z\lambda\sigma\rangle$ and $|Nk_yk_z\sigma\rangle$ respectively, where $N$ is the Landau level index, $k_y$ and $k_z$ are canonical momenta along $y$ and $z$ directions, $\sigma=\pm1$ denotes spin, and $\lambda=\pm1$ denotes the positive/negative energy states of the Dirac fermion. One may consult Refs.~\onlinecite{ShenBook, NetoRMP2009} for Landau levels of Dirac fermions, and we do not discuss the details here. After a long yet straightforward calculation of the velocity matrix elements in the Landau level states, we obtain the following expressions for the longitudinal conductivities 
\begin{align}
\sigma_{xx}^{(1)} & = \frac{e^2}{h} \frac{v_x}{v_y}\hbar^2\omega_{c1}^2 \int \frac{dk_z}{2\pi}\sum_{N\ge 0} \sum_{\lambda\lambda'}  \mathcal{M}_{\lambda\lambda'}(N,k_z)\mathcal{P}(\lambda E_{Nk_z}-\mu, \lambda' E_{(N+1)k_z}-\mu)
\label{eq:sigma1xxfiniteB} \\
\sigma_{xx}^{(2)} &  = \frac{2\kappa e^2 }{h} \sqrt{\frac{m_y^*}{m_x^*}}\hbar^2\omega_{c2}^2 \int \frac{dk_z}{2\pi}\sum_{N\ge 0}   (N+1) \mathcal{P}(E_{Nk_z}' -\mu,E_{(N+1)k_z}'-\mu) 
\label{eq:sigma2xxfiniteB}
\end{align}
where $\sigma_{xx}^{(1)}$ and $\sigma_{xx}^{(2)}$ are contributions from the Dirac and quadratic fermions respectively, and
\begin{align}
& \mathcal{M}_{\lambda\lambda'}(N,k_z)   = \frac{1}{2} -\frac{\lambda\lambda'}{2} \frac{\hbar^2v_z^2k_z^2 +m^2}{E_{Nk_z}E_{(N+1)k_z}},  \nonumber\\
&  E_{Nk_z}   = \sqrt{m^2 +\hbar^2\omega_{c1}^2 N + \hbar^2v_z^2k_z^2},\nonumber\\
& E_{Nk_z}'  =\hbar\omega_{c2}(N+1/2) + \hbar^2k_z^2/2m_z^* + \Delta
\label{eq:M}
\end{align}
The conductivities $\sigma_{yy}^{(i)}$ can be obtained by swapping the indices $x$ and $y$ in the above expressions. Calculations of the Hall conductivities $\sigma_{xy}^{(1)}$ and $\sigma_{xy}^{(2)}$ are very similar, which give rise to the following expressions
\begin{align}
\sigma_{xy}^{(1)} & = \frac{e^2}{h} \hbar^2\omega_{c1}^2 \int \frac{dk_z}{2\pi} \sum_{N\ge 0} \sum_{\lambda\lambda'} \mathcal{M}_{\lambda\lambda'}(N,k_z) \mathcal{Q}( \lambda E_{Nk_z}-\mu,\lambda' E_{(N+1)k_z}-\mu)
\label{eq:sigma1xyfiniteB}\\
\sigma_{xy}^{(2)} & = \frac{2 \kappa e^2 }{h} \hbar^2\omega_{c2}^2\int \frac{dk_z}{2\pi}\sum_{N\ge 0}   (N+1) \mathcal{Q}(E'_{Nk_z}-\mu, E'_{(N+1)k_z}-\mu) 
\label{eq:sigma2xyfiniteB}
\end{align}
Note that the function $\mathcal{Q}(E_a, E_b)$ is antisymmetric in $E_a$ and $E_b$, so the Hall conductivities are zero when $B=0$. With these expressions, we then compute the resistivity tensor $\hat\rho = \hat\sigma^{-1}$ numerically, with the chemical potential $\mu(B,T)$ determined in Eq.~\eqref{eq:density} in the main text. The numerical results are shown in Fig.~\ref{fig:rho(TB)}. One can check that in the zero-field limit $B\rightarrow 0$,  Eqs.~\eqref{eq:sigma1xxfiniteB}, \eqref{eq:sigma2xxfiniteB},  \eqref{eq:sigma1xyfiniteB} and \eqref{eq:sigma2xyfiniteB} reduce to the zero-field expressions\eqref{eq:sigmaxx1}, \eqref{eq:sigmaxx2}, \eqref{eq:sigmaxy1} and \eqref{eq:sigmaxy2}.

\begin{figure}[b]
\centering
\begin{tikzpicture}
\node at (0,0){\includegraphics[scale=0.5]{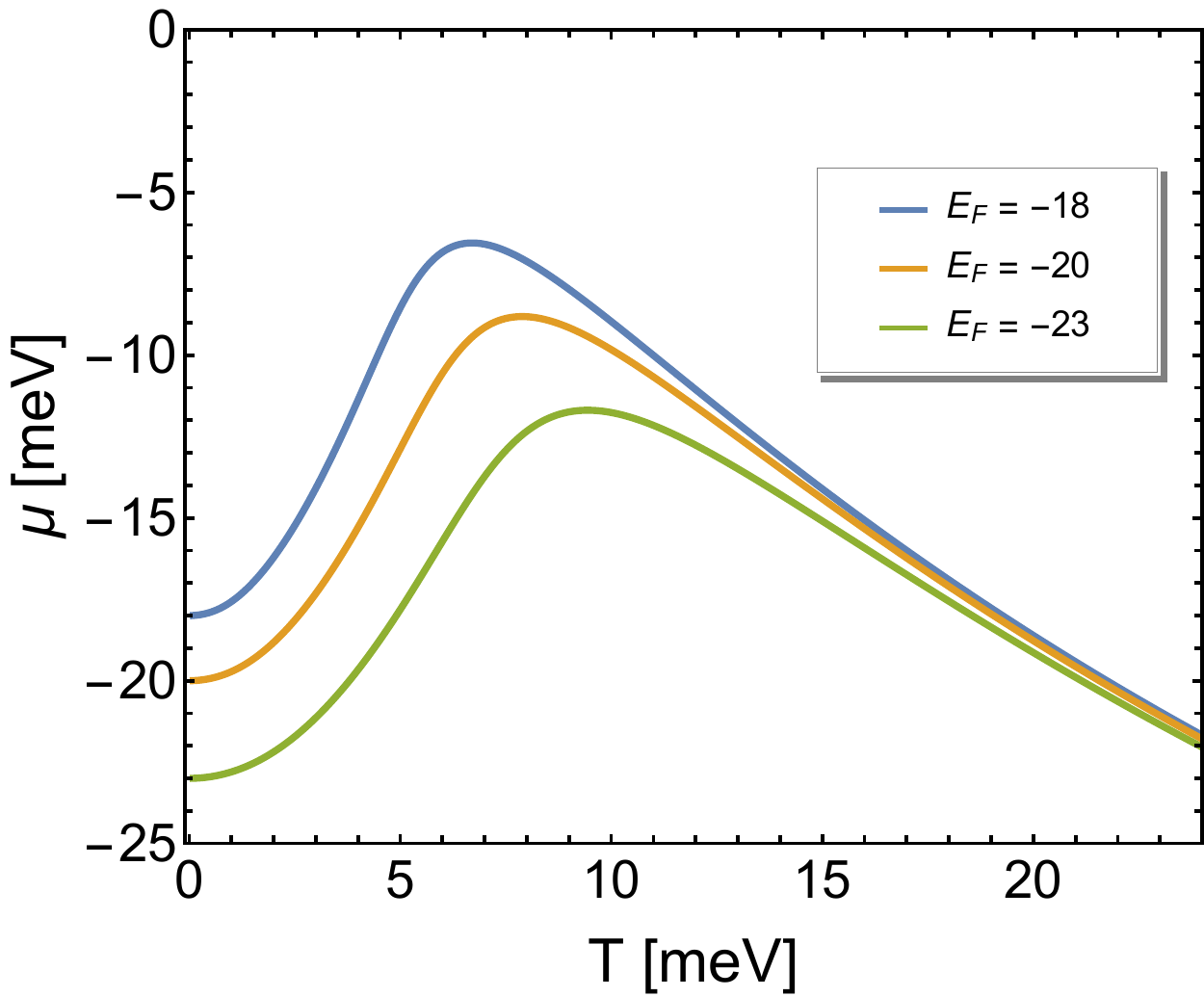}};
\node at (-2.8,3.)[scale=1.1]{\bf (a)};
\node at (8,0){\includegraphics[scale=0.55]{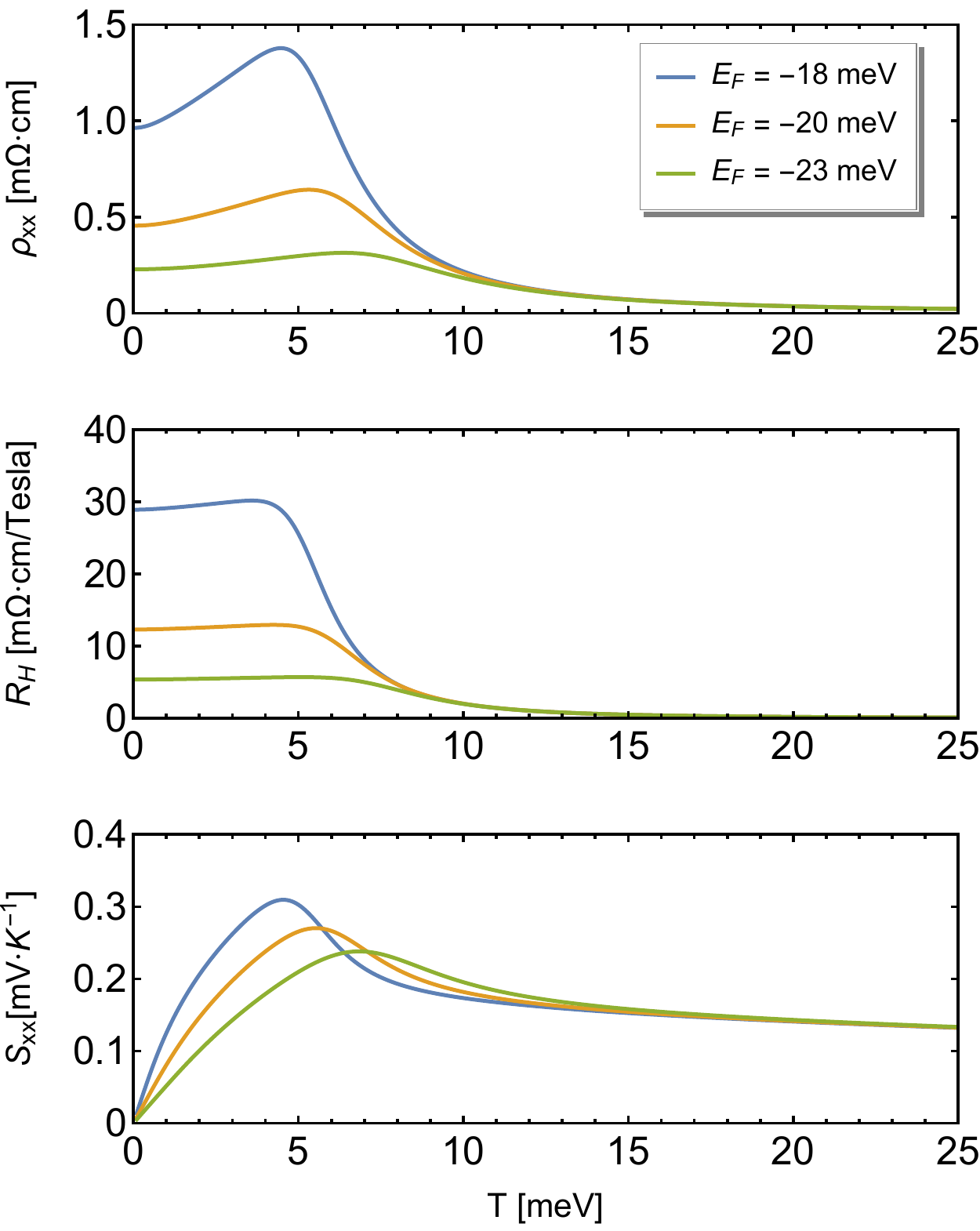}};
\node at (4.5,4.6)[scale=1.1]{\bf (b)};
\end{tikzpicture}
\caption{(a) $\mu(T)$ for negative Fermi energies at zero field. (b) Corresponding temperature dependences of $\rho_{xx}$, $R_H$ and $S_{xx}$. Numerical data are the same as those in Figs.~\ref{fig:muT} and \ref{fig:transport_B0}  of the manuscript.}
\label{fig:additiona2}
\end{figure}

\section{Additional plots for zero-field transport}

\subsection{Role of Dirac mass $m$}
\label{sec:app-massless}
In the main text, we only emphasize the importance of the energies $E_F$ and $\Delta$ for the transport anomaly. Readers may wonder if the Dirac mass $m$ plays any important role. The answer is no. At least, no significant difference is observed in our calculations between $m=0$ and $m\neq0$. In Fig.~\ref{fig:additional}(a), we have plotted $\rho_{xx}(T)$, $R_H(T)$, and $S_{xx}(T)$ in the case of massless Dirac fermion at zero magnetic field. The shapes of the curves are very similar to those in Fig.~\ref{fig:transport_B0}. Intuitively, it is expected that a non-zero $m$ becomes crucial only when the chemical potential $\mu$ moves into the band gap. However, in our model, this occurs only near the temperature $T_0$. Then, for $m \sim T_0$, the effect of Dirac mass is smoothed out by temperature and no significant difference should be observed between massive and massless Dirac fermions.

\subsection{Born approximation of disorder}
\label{sec:app-Born}
In all our calculations, we have used constant $\Gamma$ to treat the effect of disorder. To justify our simplification, here we present the result for zero-field transport using Born approximation. For simplicity, we will neglect the transport contributions from the quadratic fermions, as they are very small compared to those of the Dirac fermion. In Born approximation, the energy broadening $\Gamma(\epsilon)$ is proportional to the density of states per spin and band. Accordingly, let $\Gamma_1(\epsilon) = \gamma \alpha_1 |\epsilon|\sqrt{\epsilon^2-m^2}\Theta(\epsilon^2-m^2)$. Then, following similar steps in Appendix \ref{sec:app_transportB=0}, we find 
\begin{align}
\sigma_{xx}^{(1)} & \approx \frac{v_x^2 \hbar e^2}{3\gamma} \int_m^\infty d\epsilon \frac{\epsilon^2-m^2}{\epsilon^2} [-f_T'(\epsilon-\mu)-f_T'(-\epsilon-\mu)]\nonumber\\
\sigma_{xy}^{(1)} & \approx -\frac{e^3\hbar^2 B}{c} \frac{v_x^2 v_y^2}{6\gamma^2\alpha_1}\int_m^\infty d\epsilon \frac{\sqrt{\epsilon^2-m^2}}{\epsilon^4} [-f_T'(\epsilon-\mu) + f_T'(-\epsilon-\mu)]\nonumber\\
\varepsilon_{xx}^{(1)} & \approx  -\frac{v_x^2 \hbar e}{3T\gamma} \int_m^\infty d\epsilon  \frac{\epsilon^2-m^2}{\epsilon^2} [-(\epsilon-\mu)f_T'(\epsilon-\mu)+(\epsilon+\mu)f_T'(-\epsilon-\mu)]  
\label{eq:sigma-Born}
\end{align} 
We will set $\gamma = 3\times 10^7$ meV$^2\cdot$ \AA$^{3}$ such that $\Gamma_1(\epsilon=25\mathrm{meV})\approx 0.5$ meV. The resistivity   $\rho_{xx}(T)$, Hall coefficient $R_H(T)$ and Seebeck coefficient $S_{xx}(T)$ are plotted in Fig.~\ref{fig:additional}(b). No qualitative difference between Fig.~\ref{fig:additional}(b) and Fig.~\ref{fig:transport_B0} is observed. 

\subsection{Hole-doped systems}
\label{sec:app-hole}

In the main text, we have focused on electron-doped systems ($E_F>0$). Hole-doped samples ($E_F<0$) are also commonly seen in experiments, e.g., Refs.~\cite{ShahiPRX2018,LiangNatPhys2018,ZhangNatComm2021}. Accordingly, we have plotted $\mu(T)$, $\rho_{xx}(T)$, $R_H(T)$ and $S_{xx}(T)$ in Fig.~\ref{fig:additiona2} at zero magnetic field for a few negative Fermi energies. A few features deserve some attention:
\begin{enumerate}
\item The chemical potential $\mu(T)$ does not change sign as $T$ increases and the magnitude of its change is much smaller than the electron-doped case. The change is roughly $10$ meV, compared to $\sim 40$ meV in Fig.~\ref{fig:muT}(b). This qualitatively agrees with Ref.~\onlinecite{ZhangNatComm2021} that no or little change in $\mu$ was observed in hope-doped samples. 

\item The Hall and Seebeck coefficients remain positive in the calculated temperature region. This is expected as $\mu(T)$ does not change sign. Ref.~\cite{ShahiPRX2018} has a sample that displays a positive Seebeck coefficient throughout the measured temperature region. 

\item The longitudinal resistivity $\rho_{xx}(T)$ displays a small peak at low temperature. This peak is much lower than in the electron-doped samples: it is about 50\% larger than the residual resistivity, while in electron-doped cases the peak is about $3\sim5$ times of the residual resistivity, as shown in Fig.~\ref{fig:transport_B0}. Nevertheless, this small peak was not observed in Refs.~\cite{ShahiPRX2018,ZhangNatComm2021}. One possibility is that the peak is an artifact of our simple model. It may disappear if disorder is treated more seriously or if interaction is taken into accounts. Another possibility is that Fermi energies of the samples in Refs.~\cite{ShahiPRX2018,LiangNatPhys2018} are too close to the top of the valence band (this could be the case because the residual resistivities there are an order of magnitude larger than in typical electron-doped samples). In that case, the carrier density is so low that spatial inhomogeneity or disorder destroys the peak. If $E_F$ is further lowered in hole-doped samples, this small peak may be observed. 
\end{enumerate} 
These quantitative discrepancies suggest that more effort is needed both in experiment and theory.

\end{document}